\documentclass[preprint]{aastex}

\newcommand \eg     {{\it e.g., }}

\newcommand \pk     {{\Psi}_k}
\newcommand \ph     {\Phi}
\newcommand \Sumk   {\Sigma_k\ }
\newcommand \eq     {\,=\,}                 % separate but equal

\begin{document}

\title{Speckle Decorrelation and Dynamic Range in Speckle Noise Limited Imaging}

\author{Anand Sivaramakrishnan}
	\affil{Space Telescope Science Institute\\
		3700 San Martin Drive, Baltimore, MD 21218}
		%\email{anands@stsci.edu}
	%
\author{James P. Lloyd}
	\affil{Astronomy Department\\
		University of California
		Berkeley, CA 94720}
\author{Philip E. Hodge}
	\affil{Space Telescope Science Institute\\
		3700 San Martin Drive, Baltimore, MD 21218}
		\and
\author{Bruce A. Macintosh}
	\affil{Institute of Geophysics and Planetary Physics, \\
		Lawrence Livermore National Laboratory\\
		University of California
		Livermore, CA 94550}
\begin{abstract} %%%%%%%%%%%%%%%%%%%%%%%%%%%%%%%%%%%%%%%%%%%%%%%%
The useful dynamic range of an image in the diffraction limited regime 
is usually limited by speckles caused by residual phase 
errors in the optical system forming the image.
The technique of speckle decorrelation involves introducing many independent
realizations of  additional phase error into a wavefront
during one speckle lifetime, changing the instantaneous speckle pattern. 
A commonly held assumption is that this results in 
the  speckles being `moved around' at the rate at which the additional
phase screens are applied.
The intention of this exercise is to  smooth the
speckles out into a more uniform background distribution during their 
persistence time, thereby enabling companion detection around bright
stars to be photon noise limited rather than speckle-limited.
We demonstrate analytically why this does not occur,
and confirm this result with numerical simulations.
We show that the original speckles must persist, and that the technique of
speckle decorrelation merely adds more noise to the original
speckle noise, thereby degrading the dynamic range of the image.
\end{abstract}


%%%%%%%%%%%%%%%%%%%%%%%%%%%%%%%%%%%%%%%%%%%%%%%%%%%%%%%%%%%%%

\section{Introduction}

The recent indirect detection of extra-solar planets has fueled
extensive interest in the prospects for direct detection of light from
an extra-solar planet with either ground based adaptive optics, or a
space coronagraph.  Even moderate sized telescopes achieve sufficient
resolution to spatially resolve the planet from the parent star.
However, the challenge is achieving sufficient contrast to discriminate
the light from the planet against the residual background light 
from the star.  To the extent this background is stable, it can
be subtracted.
However, the subtraction of the point-spread function
(PSF) is typically limited by temporally variable PSF fluctuations
(speckles), which are not stable enough to be subtracted, yet
not variable enough to average out to the required level.

Ground-based adaptive optics (AO) systems correct
incoming stellar wavefronts in real-time to create a diffraction-limited
image.  Based on photon statistics for an AO corrected image, 
\citet{Nakajima94} concluded that direct detection of extrasolar
planets was feasible, even with 4 meter class ground based telescopes.
However, the well-corrected AO image is composed of a
diffraction-limited, bright core accompanied by speckles of light
which are the result of imperfect correction by the AO system.  It is the
existence and relative longevity of these residual speckles which limit the
dynamic range of long-exposure images, not photon noise \citep{Racine99}.  
Space telescopes suffer from similar effects,
variable speckles resulting from mid-frequency mirror polishing 
errors, modulated by internal spacecraft motions and vibrations.

The fundamental idea of speckle decorrelation (also known as ``phase
boiling'' or ``hyperturbulation'' \citep{Saha02}) is to scramble the
bright residual speckles of the image into a smooth background on a
faster timescale than the speckle dwell time, thereby reducing the local
spatial variance of the intensity distribution of the PSF, in order to
increase our ability to detect a faint companion or faint structure near
the star \citep{Angel94}.  This additional phase noise can be either
deliberately introduced, or is simply the result of wavefront sensing noise
in an AO system.  This concept, although widely cited
\citep{Saha02,Canales00,Boccaletti01,Racine99,Woolf98,Angel95,StahlandSandler},
has not been demonstrated to exist.  \citet{StahlandSandler} attempted to
address this problem with simulations, but these simulations were
primarily directed towards simulating the AO system, and do
not prove or disprove the existence of speckle decorrelation.  
The effect of boiling phases has also been
investigated in the context of the dark speckle method
\citep{Labeyrie95, Aime00}, with the conclusion that appropriate phase
boiling can improve SNR.  However dark speckle methods are fundamentally
different from direct imaging in that they select preferred components of
the PSF fluctuations, and dark speckle conclusions do not apply to long
exposure images.

\section{Second order expansion of the PSF}

The telescope entrance aperture and all phase effects in a monochromatic
wavefront impinging upon the optical system
can be described by a real aperture illumination function $A(x,y)$
multiplied by a unit modulus function $e^{i\phi(x,y)}$.
Aperture plane coordinates are $(x,y)$ in units of the wavelength of
the light, and image plane coordinates are $(\xi, \eta)$ in radians.
Here phase variations induced by the atmosphere or imperfect optics
are described by a real wavefront phase function $\phi$.
We neglect the effects of scintillation.
We can force $\phi$ to possess a zero mean value over the
entire aperture plane without any loss of generality, so
                        \begin{equation} \label{zeromean}
         \int A \phi \, dx dy \,\,\Big/ \int A \,dx dx \eq  0.
                        \end{equation}
Although it is not a requirement imposed by the mathematics, it is most
convenient to locate the image plane origin at the centroid of the
image PSF, which corresponds to a zero mean tilt of the wavefront
over the aperture.

At any location in  the pupil plane, the function $e^{i\phi(x,y)}$
can be expanded in an absolutely convergent series for any finite
values of the phase function:
                                                \begin{equation} \label{asf_expansion}
          e^{i\phi(x,y)} \eq   1 + i\phi  -  \phi^2/2 + ... .
                                                \end{equation}

The electric field strength in the image plane, $E(\xi,\eta)$, is the Fourier
transform (FT) of $Ae^{i\phi(x,y)}$.  Truncating the above expansion above the
second order in $\phi$, we calculate the PSF of the image to be
												\begin{eqnarray} 
												\label{expan} 
    p(\ph) &=&  a  a^*  \\
        && - i [a  (a^* \star \ph^*)  -  a^*  (a \star \ph)] \nonumber  \\
        && + (a \star \ph)  (a^* \star \ph^*)   \nonumber \\
        && -   \frac{1}{2} [a  (a^* \star \ph^* \star \ph^*)
                       + a^*  (a \star \ph \star \ph)] \nonumber ,
												\end{eqnarray}
where $a$ is the Fourier transform of the real aperture illumination function $A$,
$\Phi$ the Fourier transform of the AO-corrected wavefront phase error $\phi$,
$\star$ denotes the convolution operator and $^*$ indicates complex conjugation.  
This truncated expansion for the PSF is valid when the aberration at any point
in the pupil is significantly less than a radian.

The first term in the expansion in equation (\ref{expan}) is
the perfectly-corrected PSF $p_o = a  a^*$, which is symmetric, regardless of
aperture geometry or apodization.

The second term,
												\begin{equation} 
        p_1(\ph) =  - i [a  (a^* \star \ph^*)  -  a^*  (a \star \ph)] 
                 =  2{\rm Im} (a (a^* \star \ph^*)),
												\end{equation} 
is a real, antisymmetric perturbation of the perfect PSF.
It is modulated in size by the amplitude-spead function (ASF) $a$:
any bright first order speckle due to this term is 
`pinned' to the bright rings of the perfect ASF \citep{Bloemhof01}.
However, because $a  (a^* \star \ph^*)$ is Hermitian, 
such speckles must be accompanied by a corresponding dimming
of the PSF at a diametrically opposed point in the image.
This result is also independent of aperture geometry, though pupil
apodization will affect the underlying ring structure of $a$.

The third term,
												\begin{equation} 
         p_{2\ halo}(\ph) = (a \star \ph)  (a^* \star \ph^*),
												\end{equation} 
is real and non-negative everywhere, zero at the image center
(because of our choice of the phase zero point as described
by equation (\ref{zeromean})),
and symmetric about the center.
It is the power spectrum of the real function  $A \phi$.
This term places speckles in the dark areas of a
monochromatic PSF, and therefore sets the ultimate limits
on the dynamic range of any observational `speckle sweeping'
techniques described in \citet{Bloemhof01}.

The last term,
												\begin{equation} 
         p_{2\ Strehl}(\ph) =  -   \frac{1}{2} [a  (a^* \star \ph^* \star \ph^*) + 
		  a^*  (a \star \ph \star \ph)],
												\end{equation} 
is also modulated by the size of the ASF $a$, just like the first
order term, $p_1(\ph)$.  At the origin it reduces to the extended
Mar\'echal approximation relating the Strehl ratio $S$
to the variance of the phase over the aperture, $\sigma_{\phi}^2$,
at high Strehl ratios: 
    $S \ \simeq\  1 - \sigma_{\phi}^2$.

We analyze the statistics of  speckle contamination of
images using this second order expansion for the PSF of a well-corrected image.

\section{The speckle decorrelation model}

The speckle decorrelation technique can be modelled by
adding $N$ uncorrelated, independent
realizations of phase noise $\psi_k(x)$ to the wavefront
(with a deformable mirror, for example) while the exposure
is in progress.  The index $k$ runs from $1$ through $N$.
%Many independent realizations of phase noise are added to the
%wavefront during the lifetime of a single speckle.
We assume that the $N$ artificial phase error realizations are
asserted for equal time intervals, $\tau = T/N$, during a
speckle lifetime $T$.  Each of the $\psi_k$'s is constructed to
possess a zero aperture-weighted mean, as well as
the same mean tilt across the aperture as $\phi$.

The PSF of the exposure which lasts the length of 
the speckle lifetime $T$  will therefore be the average of
individual PSF's ${p_k}$, each of which is formed by a wavefront with
a phase of $(\phi + \psi_k)$.
The PSF during the time the $k^{th}$
phase screen is added to the residual phase $\phi$ is
												\begin{equation} 
												\label{pk} 
	p_k \eq p(\ph+\pk), 
												\end{equation} 
where $\pk$ is the Fourier transform of $\psi_k$ (using equation (\ref{expan})).
The PSF of an image with exposure time T is the average of each of the
individual PSF's: 
												\begin{equation} 
												\label{avgpsf} 
		\bar{p} = \Sumk p_k / N.
												\end{equation} 
Since $a$ and $\ph$ are constant for the duration of the exposure,
the summation in equation (\ref{avgpsf}) can be moved to within the
multiplications and convolution integrals in equation (\ref{pk})
and (\ref{expan}).

%% This is obviously the original speckled halo with a smooth background
%% with an intensity which is the same as the local (in image plane location)
%% variance as the ensemble of speckle patterns (if the set of `decorrelating'
%% phase screens ${\psi_k}$ was chosen to match the post-AO atmospheric
%% statistics).  All this does is to fill the gaps between speckles 
%% with unwanted light, which {\it reduces} the dynamic range of the image in
%% the case of photon statistics-limited backgrounds.
%% 

If we expand $p_k$ in the manner of equation (\ref{expan}), we obtain
												\begin{eqnarray} 
												\label{pkexpan} 
    {p_o}_k &=&  p_o(\ph)  \nonumber  \\
    {p_1}_k &=&  p_1(\ph) + p_1(\pk) \nonumber  \\
    {p_{2\ halo}}_k &=&  p_{2\ halo}(\ph) + p_{2\ halo}(\pk)  \nonumber \\
				&&	+ (a \star \ph)  (a^* \star \pk^*)  
               		+ (a^* \star \ph^*)  (a \star \pk)  \nonumber   \\
    {p_{2\ Strehl}}_k &=&   p_{2\ Strehl}(\ph) + p_{2\ Strehl}(\pk) -   \nonumber  \\
				&& [a   (a^* \star \ph^* \star \pk^*)  
				  + a^* (a   \star \ph   \star \pk)].  \nonumber   
												\end{eqnarray}
Averaging each of these terms over the N realizations of $\psi_k$,
with their corresponding transforms $\pk$, produces average intensities
of 
												\begin{eqnarray} 
												\label{pkavg} 
    {\bar{p}_o} &=&  p_o(\ph)  \nonumber  \\
    {\bar{p}_1} &=&  p_1(\ph) + \Sumk p_1(\pk)/N \nonumber  \\
    {\bar{p}_{2\ halo}} &=&  p_{2\ halo}(\ph) + \Sumk p_{2\ halo}(\pk) / N \nonumber \\
				&+&	 (a \star \ph)  (a^* \star \Sumk \pk^*) / N      \nonumber \\
               	&+&	 (a^* \star \ph^*)  (a \star \Sumk \pk)  / N     \nonumber \\
    {\bar{p}_{2\ Strehl}} &=&   p_{2\ Strehl}(\ph)                       \nonumber \\
	                &+&   \Sumk p_{2\ Strehl}(\pk) / N               \nonumber \\
				&-&  [a   (a^* \star \ph^* \star \Sumk \pk^*)        \nonumber \\
				&+&   a^* (a   \star \ph   \star \Sumk \pk)] / N   \nonumber   
												\end{eqnarray}
in the final image.  The sum of these individual averages is
the final PSF of the image.  
The quantity $S_k(\xi,\eta) = \Sumk \pk$ or its conjugate appear frequently
in the averaged PSF.  $S_k$ is zero mean, with a  standard 
deviation of $\sigma_\Psi / \sqrt{N}$ (where $\sigma_\Psi^2(\xi,\eta)$
is the variance of the parent distribution of the random phase functions'
Fourier transforms).

The zeroth order term remains $a a^*$.

The first order contribution of the $N$ `speckle decorrelating'
phase screens to $\bar{p}$ is $\Sumk p_1(\pk)/N$,
which is a zero-mean quantity.  It can be rewritten as
$2 {\rm Im}(a (a \star S_k)) $.

The $\bar{p}_{2\ halo}$ term is composed of the sum of the original halo term
and the expectation value of the halo term calculated over the ensemble of
phase error functions, because both
	 $(a \star \ph)  (a^* \star S_k^*) / N$
and
     $(a^* \star \ph^*)  (a \star S_k) / N$
have zero expectation values.

Likewise, $\bar{p}_{2\ Strehl}$ term is composed of the sum of the original Strehl term
and the expectation value of the individual Strehl term calculated over the ensemble
of the phase error functions, because
$  a   (a^* \star \ph^* \star S_k^*)$  and  
$  a^* (a   \star \ph   \star S_k)  $ have zero expectation values.

None of the zero-mean terms contribute in a manner that will alter the long
exposure image in the limit as $N$ becomes large.  Therefore, the
resulting image formed by the addition of small (in the perturbation
theory sense) phase noise does not result in speckle decorrelation.

%%%%%%%%%%%%%%%%%%%%%%%%%%%%%%%%%%% ver 2 bmac 
\section{Numerical simulations}

We have carried out numerical simulations to illustrate this point. Figure
1 (a) shows a simulated image created from a small static phase error with a
RMS magnitude of 17 nm at a wavelength of 1.6 microns.
(In this case, the image simulated the residual 
atmospheric fitting error in an adaptive optics correction, but we could have
used any
any error with an approximately flat power spectrum over the region of
the image.)  Diffraction rings have
been suppressed at radii $> 0.1$" by using pupil apodization.
We then simulated a random phase error 
by injecting 20 nm of 
white phase noise into the phase residuals. These were convolved with a Gaussian 
kernel to represent the same actuator spacing as was used to generate the 
fitting error (following the method outlined in \citet{Sivaramakrishnan01}),
but this
could represent any error with a similar power spectrum to the static error. 
Figure 1 (b) shows the resultant image, showing an entirely new speckle 
pattern. We simulated the long-exposure image process by keeping the 
static error fixed, injecting different realizations of 
the white noise, and adding together the
resultant images, thereby attempting to smooth out
a long lived speckle pattern with a very fast-moving speckle
pattern. Figure 1 (c) shows the result after 20 iterations, and Figure
1 (d) after 100 iterations; the image speckle pattern has returned to the
original pattern, offset by a pedestal equal to the average intensity
of the random speckle pattern. 

\begin{figure}[htbp]
\epsscale{0.55}
      \plotone{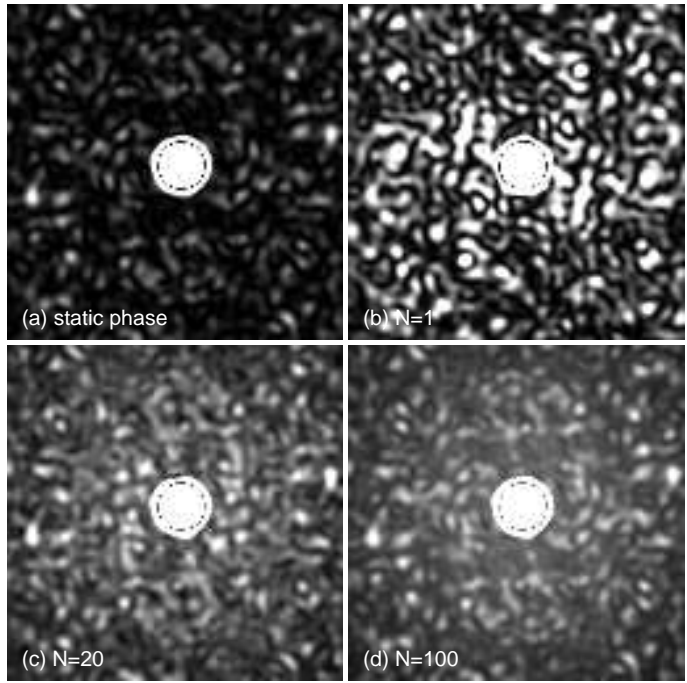}
      \caption{
		(a) A simulated (\eg\  AO-corrected) PSF with a residual
		fitting error of 17 nm rms, producing a speckle pattern.
		(b) The PSF of a 20 nm rms independent phase error added to that 
		used in (a).
		(c) the sum of 20 PSFs, each with the same residual error as (a) and 
		independent 20 nm phase errors as in (b).
		(d) the sum of 100 PSFs as in (c)   
		(See text for details).
		}
		\label{fig1}
  \end{figure}
 
Figure 2 shows this quantitatively. This is a plot of the image noise
(measured as the standard deviation of the intensity values in an 
annulus) as a function of the number of seperate images added together.
The noise initially declines rapidly as the speckles due to the white noise
average together, but ultimately the noise reaches a plateau equal to 
the noise in an image having only the original fitting error common to all
the images.

  \begin{figure}[htbp]
      %\figurenum{text}
      %\epsscale{num}
\epsscale{1}
      \plotone{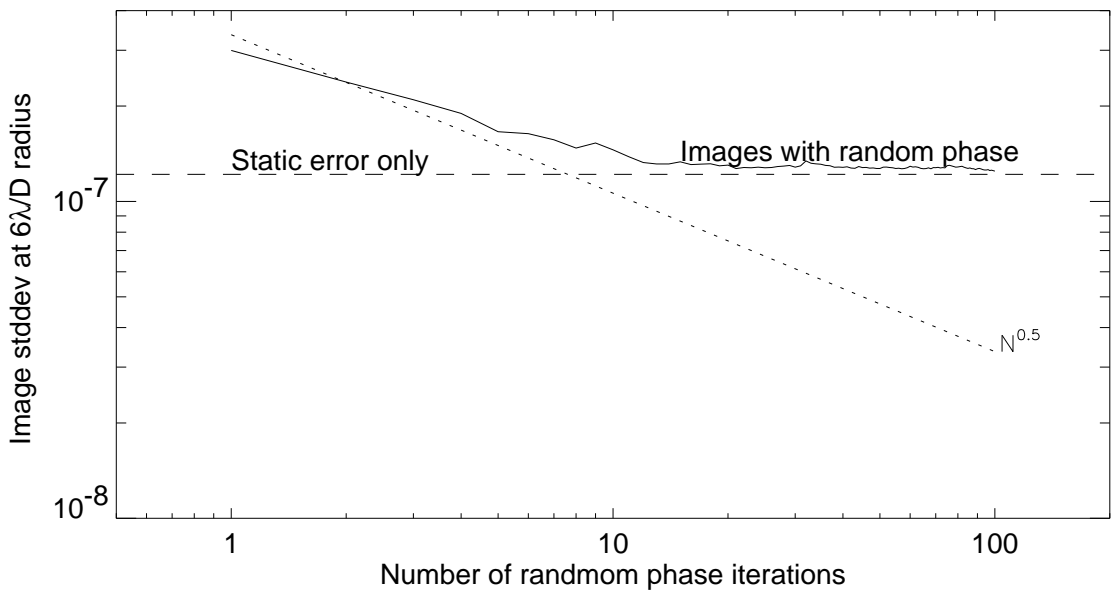}
      \caption{
Image noise (measured as the standard deviation of the intensity 
values in an annulus) as a function of the number of seperate images
added together.  The noise eventually reaches a plateau equal to the
noise in an image having only the original residual phase error. 
(See text for details).
		}
		\label{fig2}
  \end{figure}

On the face of it this appears to contradict the simulations of 
\citet{StahlandSandler}.
%Stahl and Sandler (1995). 
However, the results are actually consistent with
their work. Their initial simulations showed a long-lived speckle pattern
that changed only on atmospheric timescales, which they attributed to a 
large wavefront error caused by time lags. (It may have been augmented by 
the lack of a wavefront reconstructor algorithm --- instead, direct average
phase measurements were applied directly to their simulated DM.)  
When they changed their AO model to incorporate a predictive controller with 
lower time lags, the speckle pattern began to change rapidly. The effect of 
the predictive controller,
however, is not to decorrelate the speckles due to atmospheric timelag
but to reduce the residual wavefront error due to this source and hence
the total intensity of the long-lived speckle pattern; this left the 
short-lived speckles due to measurement error as the main remaining noise
source. Stahl and Sandler did not demonstrate speckle decorrelation, but
speckle suppression. One can always reduce speckle noise by reducing the
corresponding wavefront error term; what one cannot do is change the
timescale of a given speckle error term by introducing other zero-mean
phase errors uncorrelated with the phase errors causing the original
speckle. (In addition, the short timescales simulated by Stahl and Sandler
may have masked longer-lived speckle effects, and the coarse sampling
of their simulations may also have masked the real evolution of noise
sources.)

\section{Alternatives to speckle decorrelation}
 
Each term in the series expansion of the PSF possesses distinct properties.
Further work on characterizing the magnitude of the various terms
(in both monochromatic and polychromatic cases), assuming either
AO-corrected atmospheric turbulence or typical mirror aberrations
is being done in order to assess how one might use
the knowledge of these properties to improve dynamic range.
The first order term has already been treated in the literature
\citep{Bloemhof01, Boccaletti02},
although the second degree term $p_{2\ Strehl}$ is often the
largest term close to the image core.  Furthermore, when using 
speckle sweeping techniques, it is likely that
ultimate dynamic range limits in the wings of the PSF are set by the
other second degree term, $p_{2\ halo}$.

\acknowledgements

The authors wish to thank the Space Telescope Science Institute's
Research Programs Office and its Director's Discretionary Research Fund.
This work has also been supported by the National Science Foundation
Science and Technology Center for Adaptive Optics, managed by the
University of California at Santa Cruz under cooperative
agreement No.  AST - 9876783,
the Australian Fulbright Commission, and
the SPIE Irving J. Spiro fellowship.
Portions of this work were performed under the auspices of the U.S. Department of Energy,
National Nuclear Security Administration by the University of California,
Lawrence Livermore National Laboratory under contract No. W-7405-Eng-48.

\bibliographystyle{apj}
\bibliography{ms}

\end{document}